\documentclass[useAMS,fleqn,usenatbib]{mnras}

\usepackage[T1]{fontenc}
\usepackage{ae,aecompl}

\usepackage{newtxtext,newtxmath}
\usepackage{graphicx}
\usepackage{amssymb}
\usepackage{amsmath}
\input psfig.sty       
\RequirePackage{xcolor}

\def \grss {GRS 1915+105~}
\def \grs {GRS 1915+105}

\def \saxx {{\it Beppo}SAX~}

\defcitealias{Massaro2020}{Paper~I}

\title[A non-linear model for the variability classes of GRS~1915+105 - II]
{A non-linear mathematical model for the X-ray variability classes of the microquasar 
GRS~1915+105 - II: transition and swaying classes}  
\author[E. Massaro, F. Capitanio, M. Feroci, T. Mineo, A. Ardito, P. Ricciardi]
  {E. Massaro$^{1}$,
  F. Capitanio$^{1}$, 
  M. Feroci$^{1}$,
  T. Mineo$^{2}$\thanks{E-mail address: \texttt{teresa.mineo@ifc.inaf.it}},
  A. Ardito$^{3}$, 
  P. Ricciardi$^{3}$ \\
$^1$ INAF, IAPS, via del Fosso del Cavaliere 100, I-00113 Roma, Italy \\
$^2$ INAF, IASF Palermo, via U. La Malfa 153, I-90146 Palermo, Italy \\
$^3$ Sapienza Universit\`a di Roma, Piazzale Aldo Moro, 5, I-00185 Roma, Italy \\
}
      
\date{}

\pubyear{2020}

\begin{document}
\label{firstpage}
\pagerange{\pageref{firstpage}--\pageref{lastpage}}
\maketitle

\begin{abstract}
The complex time evolution in the X-ray light curves of the peculiar black 
hole binary \grss  can be obtained as solutions of a non-linear system of 
ordinary differential equations derived form the Hindmarsch-Rose model and 
modified introducing an input function depending on time.
In the first paper,assuming a constant input with a superposed white noise, we 
reproduced light curves of the classes $\rho$, $\chi$, and $\delta$.
We use this mathematical model to reproduce light curves, including some interesting
details, of other eight \grss variability classes either considering a variable 
input function or with small changes of the equation parameters.
On the basis of this extended model and its equilibrium states, we can arrange most 
of the classes in three main types:
$i$) {\it stable equilibrium patterns} (classes $\phi$, $\chi$, $\alpha''$, $\theta$, 
$\xi$, and $\omega$) whose light curve modulation follows the same time scale of the 
input function, because changes occur around stable equilibrium points; 
$ii$) {\it unstable equilibrium patterns} characterised by series of spikes (class
$\rho$) originated by a limit cycle around an unstable equilibrium point;
$iii$) {\it transition pattern} (classes  $\delta$, $\gamma$, $\lambda$, $\kappa$ and 
$\alpha'$), in which random changes of the input function induce transitions from stable 
to unstable regions originating either slow changes or spiking, and the occurrence 
of dips and red noise. \\
We present a possible physical interpretation of the model based on the similarity
between an equilibrium curve and literature results obtained by numerical integrations 
of a slim disc equations.
\end{abstract}
	
\begin{keywords}  stars: binaries: close - stars: individual: GRS~1915+105 -
 X-rays: stars - black hole physics
\end{keywords}

\maketitle

\section{Introduction}
 
In a previous paper \citep[][hereafter Paper I]{Massaro2020} we have shown
that the complex time evolution of some variability classes exhibited by
the peculiar Black Hole  binary (BHB) \grss are obtained as solutions of
a system of ordinary differential equations (ODEs).
It is known that light curves of \grss in the X-ray band were classified in
14 different types, and it is possible that other new types can be introduced.
The first classification was proposed by \citet{Belloni2000} who defined 12 
variability classes on the basis of a large collection of multi-epoch RXTE 
observations.
Two more classes were discovered in the following years 
\citep{KleinWolt2002, Hannikainen2003, Hannikainen2005}

The ODE system proposed in \citetalias{Massaro2020} is based on the so-called 
Hindmarsh-Rose model \citep{Hindmarsh1984, Hindmarsh2005}, widely used in the 
description of neuronal behaviour.
We introduced some changes with respect to the original formulation and refer 
to the new system as Modified Hindmarsh-Rose (MHR) model, that is non-autonomous 
because of the presence of an {\it input function} depending upon the time.
We were thus able, assuming a constant input with a superposed white noise,
to reproduce light curves of the classes $\phi$, $\chi$ and $\delta$, the last 
one characterized by a red noise Power Density Spectrum (PDS), and in particular
those of the $\rho$ class limit cycle with its quasi regular series of spikes.
An interesting finding was that the MHR model leads naturally to the onset of
low frequency QPOs when the values of the input function vary in a transition 
range between unstable and stable equilibrium.
As we wrote in \citetalias{Massaro2020} this mathematical model should be 
considered as a useful tool for describing in a unified picture some non-linear 
effects occurring in the variability classes and their transitions.

In the present paper we will show that the MHR model can be used for reproducing
X-ray light curves of several other variability classes of \grss either with an 
assumption of a variable input function or with small changes of the parameters.
The success of this mathematical model is due to the fact that it appears a quite 
good analytical approximation of some instability conditions that can occur in 
accretion discs.

In Sections~\ref{sect:2} and ~\ref{sect:3} we summarize the MHR model and extend 
it to the case of a new free parameter and a variable input function; 
in Section ~\ref{sect:4} show as this extended MHR model can account for light 
curves of the $\alpha'$, $\gamma$ and $\kappa$ classes;
in Section ~\ref{sect:5} we reproduce light curves of other variability classes 
which require rather {\it ad hoc} input functions, but, nevertheless, exhibit some 
minor interesting features; finally, in Section ~\ref{sect:6} we discuss some possible 
physical interpretation of the MHR model on the basis of some literature results 
obtained by numerical integrations of disc equations.

\section{The MHR non-linear ODE system}
\label{sect:2}

The MHR model introduced in  \citetalias{Massaro2020} consists of two ODEs:

\begin{eqnarray}
\frac{dx}{dt} &=& - \rho x^3 + \beta_1 x^2 + y + J(t) \nonumber \\
\frac{dy}{dt} &=& - \beta_2 x^2 - y  
\label{eq1}
\end{eqnarray}

In \citetalias{Massaro2020} the parameters where kept fixed to the values and 
$\beta_1 = \beta_2 = 3.0$ and the input function $J(t)$ was taken in the form

\begin{equation}
   J(t) = J_0 + C~ r
\label{eq2}
\end{equation}

\noindent
where $J_0$ is a constant, $r$ is a random number with a uniform distribution in the 
interval [$-0.5$, $0.5$], and a resulting standard deviation $\sigma_J = C/(2\sqrt{3})$ 
independent of the $J_0$ values, and $C$ is a constant factor to vary the amplitude 
of the random fluctuations.

In the present work we fixed again the value of $\rho = 1.0$, without any loss of
generality, because this is not an interesting parameter that can be eliminated by 
a simple change of variables as shown in \citetalias{Massaro2020}, while in a few 
cases $\beta_1$ and $\beta_2$ were different; moreover we substitute $J_0$ with a 
variable $J_S(t)$, whose particular shape was adapted to reproduce light curves of 
the various classes.
Thus, in place of Eq. (2) we write:

\begin{equation}
   J(t) = J_S(t) + C~ r
\label{eq3}
\end{equation}

Numerical computations were again performed by means of a Runge-Kutta fourth order 
integration routine \citep{Press2007}, that was verified very stable and fast.

\begin{figure}
\includegraphics[height=7.9cm,angle=-90,scale=1.0]{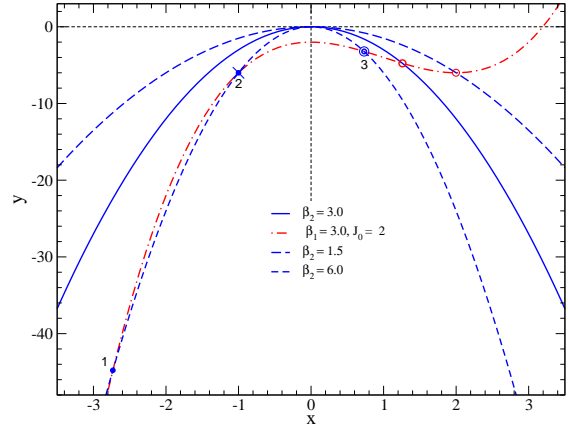}
\caption[]{
{\it Nullclines} for the system of Eq.~\ref{eq4} for $J_0 =2.0$ and $\beta_1 = 3.0$ (red 
curve); the blue curves are the parabolas given by the second of Eq.~\ref{eq4} for
different values of $\beta_2$.
The open red circles are the equilibrium points of the two low $\beta_2$ parabolae,
and the blue circles with numbers are the three equilibrium points for the high
$\beta_2$ parabola.
The cross and the large circle mark the saddle and unstable point, respectively.
}
\label{fncl_1}
\end{figure}

\section{Nullclines, equilibrium states and stability}
\label{sect:3}

In the case $J_S(t) = J_0$, the equilibrium conditions for the system of
Eq.~\ref{eq1} with $\rho = 1$, i.e. $\dot{x} = \dot{y} = 0$, are

\begin{eqnarray}
 y &=&  x^3 - \beta_1 ~ x^2 - J_0 \nonumber \\
 y &=& - \beta_2~ x^2   
 \label{eq4}
\end{eqnarray}

\noindent
thus the equilibrium points are the solutions of a cubic equation, as
explained in the Appendix~\ref{appendix1}.
An interesting possibility is that of having three equilibrium points 
instead of only one, as in \citetalias{Massaro2020}.
A lower or a higher value of $\beta_2$  corresponds to a {\it nullcline} with a
lower or a higher curvature, respectively.
Fig.~\ref{fncl_1} shows some {\it nullclines} in the $x, y$ plane: we plotted only 
one cubic {\it nullcline} with $\beta_1 = 3.0$ $J_0 = 2.0$, typical values adopted 
in our computations, together with three parabolic {\it nullclines}.
The solutions and the stability conditions are given in the Appendix~\ref{appendix1}: 
the condition to have three equilibrium points is that $\beta_2 > 5.38$.
In Fig.~\ref{fncl_1} we plotted two parabolas for low $\beta_2$ values with only one
equilibrium point, while the third one has three intersections with the cubic one, 
marked by 1, 2, and 3.
Using the linear analysis in the Appendix~\ref{appendix1} it is easy to verify that: 
$i$) for $x_{*1}$ one has $\Delta > 0$ and $Tr < 0$ and the equilibrium is stable, 
$ii$) for $x_{*2}$, $\Delta < 0$ and $Tr < 0$ corresponding to a saddle point, and
$iii$) for $x_{*3}$, $\Delta > 0$ and the equilibrium can be either stable
or unstable.
The saddle point drives trajectories along the {\it nullclines} and prevents those 
moving away from them.
It is remarkable that in the interval between $x_{*1}$ and $x_{*2}$ the two 
{\it nullclines} are very close to each other, a property relevant to understand 
the bursting process.

\subsection{Variable $J_S(t)$}
\label{sect:3p1}

Several classes are characterised by flux variations on time scales longer than those 
of the $\rho$ self-oscillations.
Such slow time scales are not intrinsic to the MHR model of Eq.~\ref{eq1} and can only 
be explained by means of a variable input function having values generally lower 
than the one necessary to develop the spiking behaviour.
In this way the solutions of Eq.~\ref{eq1} are substantially driven by $J_S(t)$, that 
can be assumed to have rather simple profiles like a step (or multistep) function or 
a power law triangular profile, with an amplitude varying between 0 and 1:

\begin{eqnarray}
 J_S(t) &=& J_m \Big[1-\Big(1-\frac{t}{t_1}\Big)^m\Big]  ~~~~, ~~~ 0 < t < t_1  \nonumber \\
 J_S(t) &=& J_m \Big(\frac{1-t/T}{1-t_1/T}\Big)^n  ~~~~~~~, ~~~ t_1 < t < T
\end{eqnarray}

\noindent
where $T$ is the duration (or period) of the modulation and $t_1$ is the time of 
the maximum amplitude $J_m$; $m$ and $n$ are real exponents that make curve the
rising and decaying sections of this function ($m = n = 1$ result in straight
lines while for very high values they approximate well a square wave pattern).
When it would be useful local and irregular features will also be considered to 
reproduce some details of the observed light curves.

\section{Solutions with $\beta_2 \ne \beta_1$}
\label{sect:4}

We verified that solutions of MHR model reproduce the light curves of some 
other classes with a satisfying accuracy if the condition $\beta_1 = \beta_2$ 
is released.
We decided, therefore, to keep $\beta_1$ fixed at 3.0, as in \citetalias{Massaro2020}, 
and change only $\beta_2$. 
In particular, as shown in the following, the condition $\beta_2 < \beta_1$ gives 
good curves for the $\kappa$ and $\gamma$ class, while $\beta_2 > \beta_1$ is
required for the $\alpha'$ class.

We point out first that the criteria for the light curve classification of \grss 
used by \citet{Belloni2000} are not completely unambiguous and, looking at the 
individual light curves listed in that paper, one can notice different structures 
The classes $\alpha$, $\nu$ and $\beta$ exhibit similar patterns characterized 
by spike bursting patterns alternating with a smooth decline and a slower rise.
The typical duration of the bursting phase is around 500 s, while the smooth 
interburst segments can be of the order of 1000 s and even longer.
In the following, therefore, we distinguish two subclasses, here indicated 
as $\alpha'$ and $\alpha''$ because their light curves have different structures
despite being reported in the same class.
According to the previous modelling the slow modulation is explained by
variations of the input function $J(t)$, while the bursting is produced by
the relaxation oscillation when the high value of $J$ leads the system to
the unstable region.

\begin{figure}
\includegraphics[height=7.9cm,angle=-90,scale=1.0]{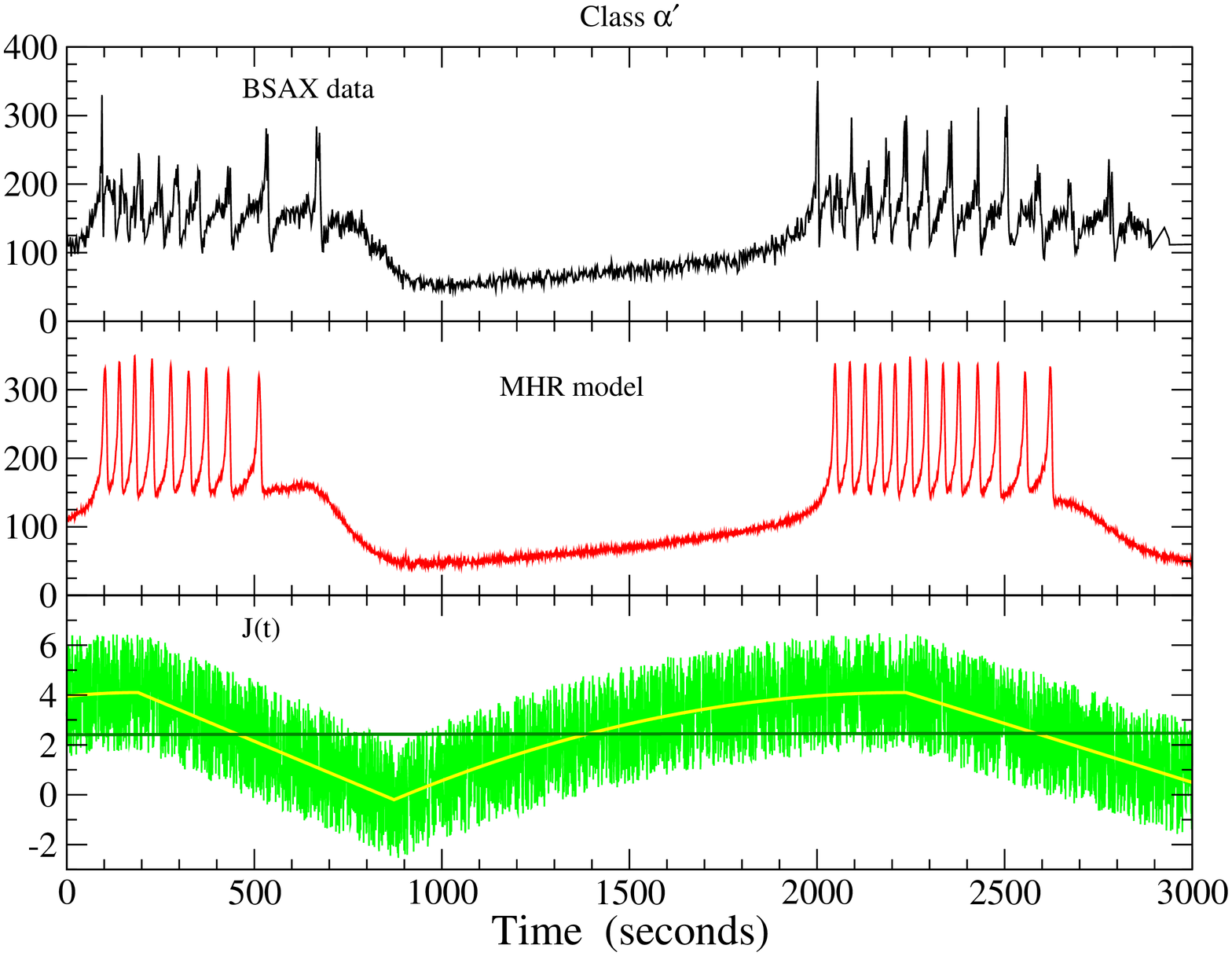}
\includegraphics[height=7.9cm,angle=-90,scale=1.0]{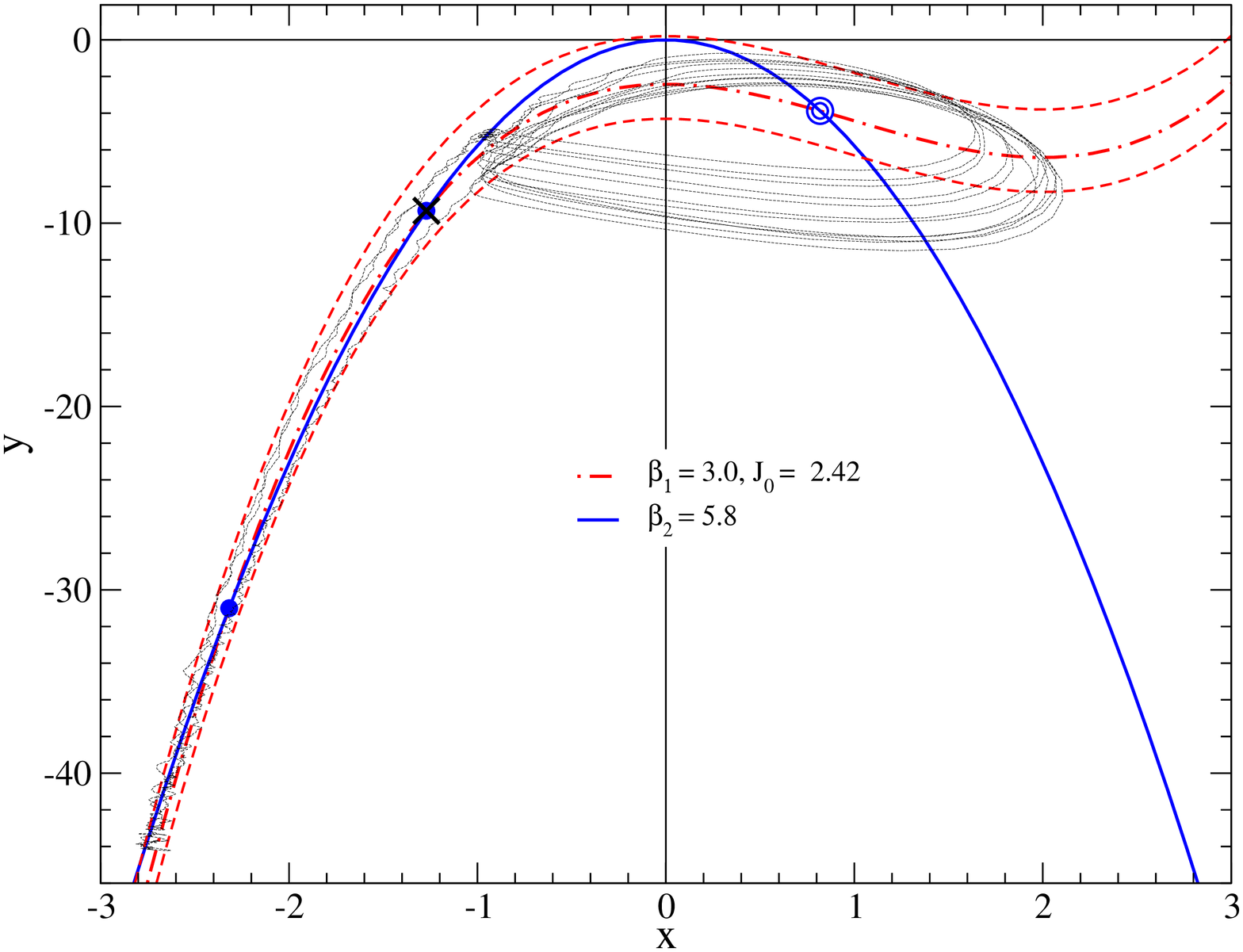}
\includegraphics[height=7.9cm,angle=-90,scale=1.0]{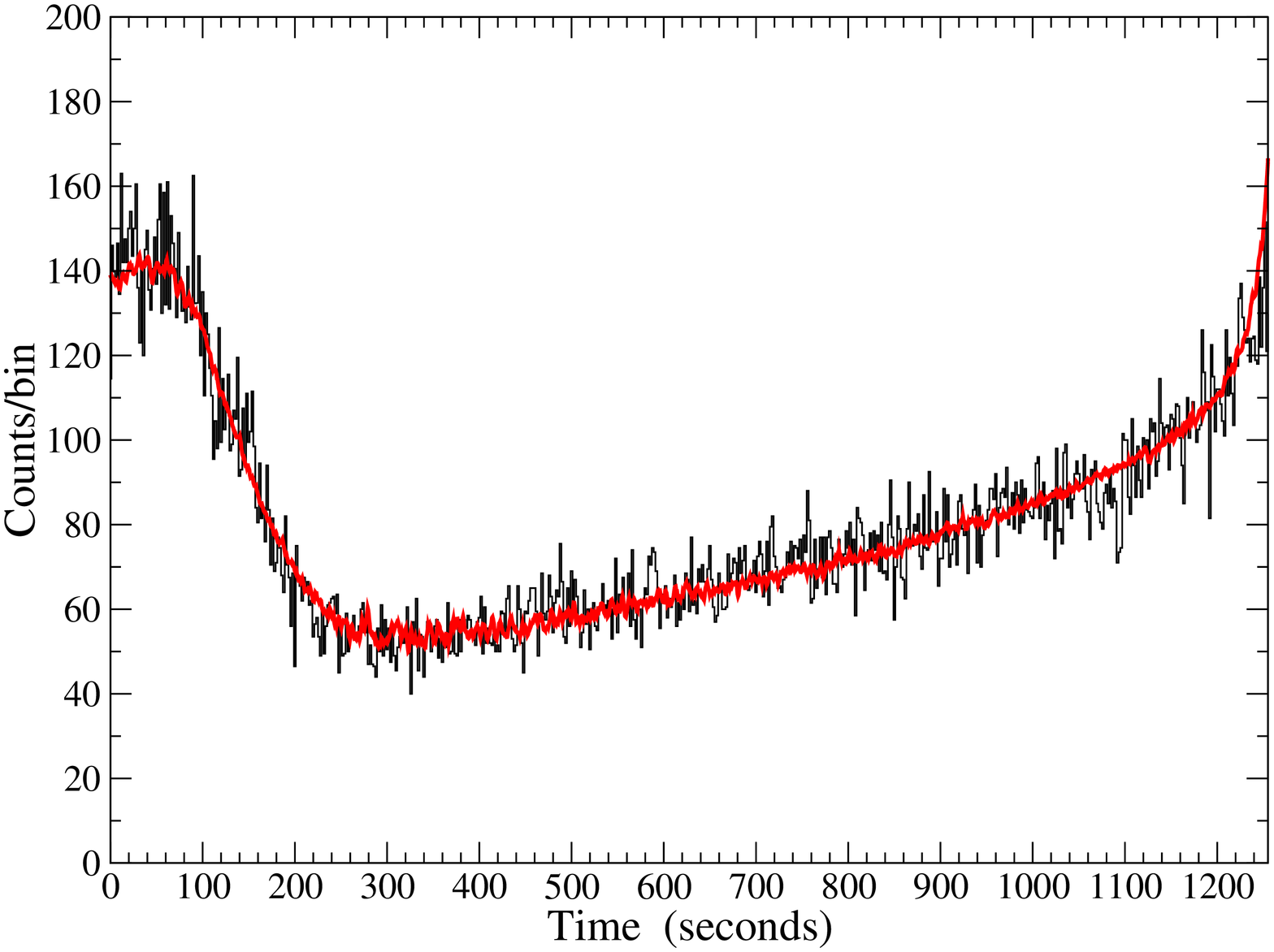}
\caption[]{
Top panel: top: short segment of MECS light curve of the $\alpha'$ class;
centre: a light curve (red) computed using the MHR model with the parameters' 
values in Table 1;
bottom: the input function $J(t)$ (green) with random fluctuations 
superposed onto a slowly variable modulation $J_S(t)$ (yellow); the dark green 
line is the mean value.\\
Central panel: the {\it nullclines} and the phase space trajectory (black) for 
the $\alpha'$ class MHR solution; the thick red curve corresponds to the
mean value of $J(t)$ and dashed ones are for the maximum and minimum values of
$J_S(t)$; \\
Bottom panel: detail of the interburst segment with the model (red) superposed 
onto data (black); the time scale was adjusted to match the total duration.
}
\label{falpr}
\end{figure}

\subsection{$\beta_2 > \beta_1$: class $\alpha'$}

An example of the $\alpha'$ class light curve, observed with the MECS \citep{Boella1997} 
on board the \saxx satellite on 1995 November 11 in the energy range 1.5-10 keV 
is shown in the top panel in Fig. ~\ref{falpr}.
It presents a short series of spikes, some of them similar to those of the $\rho_d$ 
class, with an increasing time separation between them; after the last spike the 
count rate decreases smoothly to the minimum level in about 200 s to increase again 
to a level at which the spiking activity starts again.
We were able to reproduce the most characterizing features of this pattern (top panel 
centre) by assuming a value of $\beta_2$ above the three equilibrium points threshold 
(see Appendix~\ref{appendix1}) together with a slowly modulated input function 
having a pattern resembling a sinusoidal oscillation (top panel bottom).
For low values of $J_S(t)$ the system is in the stable interval and the solutions 
are mainly determined by the shape of the {\it nullclines}.
When the value of $J_S(t)$ becomes high enough to move the system towards the unstable
region the spikes appear and the subsequent decline is responsible
of their increasing recurrence time; spikes cease when input function moves again to 
the stable region. 
It is worth noticing that the spiking behaviour is a consequence only of the
system entering the unstable region and it is not specifically triggered by features 
in the $J(t)$ function.

\begin{table}
\caption{Parameters' values adopted in the numerical calculations of the light curves 
for the various variability classes of \grs.}
{\small
\begin{tabular}{crrrrrrr}
\hline
   Class     & $\beta_1$ & $\beta_2$ &  $C$  &  $J_0$  &   $J_m$  & $\langle J \rangle$ &  \\
\hline
             &           &           &       &         &          &                     &  \\ 
 $\alpha'$   &      3.0  &      5.8  &  4.00 &  $-$0.2 &   4.30   &        2.420        &  \\ 
             &           &           &       &         &          &                     &  \\ 
 $\gamma$    &      3.0  &      2.1  &  7.00 &    2.50 &   2.40   &        3.704        &  \\ 
 $\kappa$    &      3.0  &      1.6  &  6.00 & $-$2.20 &   6.60   &        1.859        &  \\
             &           &           &       &         &          &                     &  \\ 
 $\lambda$   &      4.0  &      4.0  &  5.00 &    1.45 &   0.00   &      $-$2.515       &  \\ 
             &           &           &       &         &          &                     &  \\ 
 $\omega$    &      3.0  &      3.0  &  4.50 & $-$2.30 &   1.20   &      $-$0.348       &  \\ 
 $\alpha''$  &      3.0  &      3.0  &  10.5 & $-$6.70 &   1.30   &      $-$0.797       &  \\ 
 $\xi$       &      3.0  &      3.0  &  7.00 & $-$2.70 &   2.81   &      $-$1.137       &  \\ 
 $\theta$    &      3.0  &      3.0  &  4.50 & $-$3.20 &   2.20   &      $-$2.475       &  \\ 
\hline

\end{tabular}
}
\label{tab:param}
\end{table}

\begin{figure}
\includegraphics[height=7.9cm,angle=-90,scale=1.0]{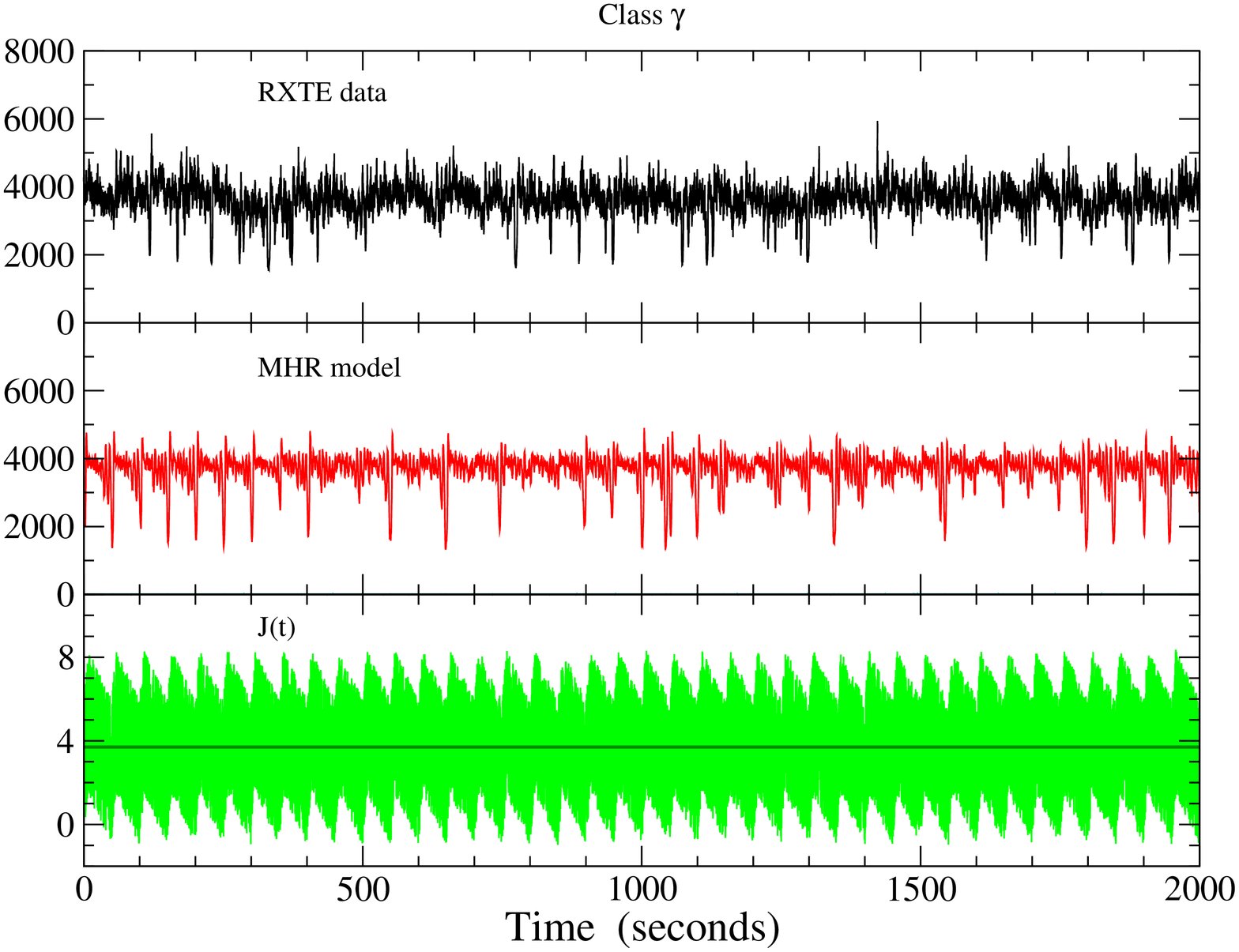}\\
\includegraphics[height=7.9cm,angle=-90,scale=1.0]{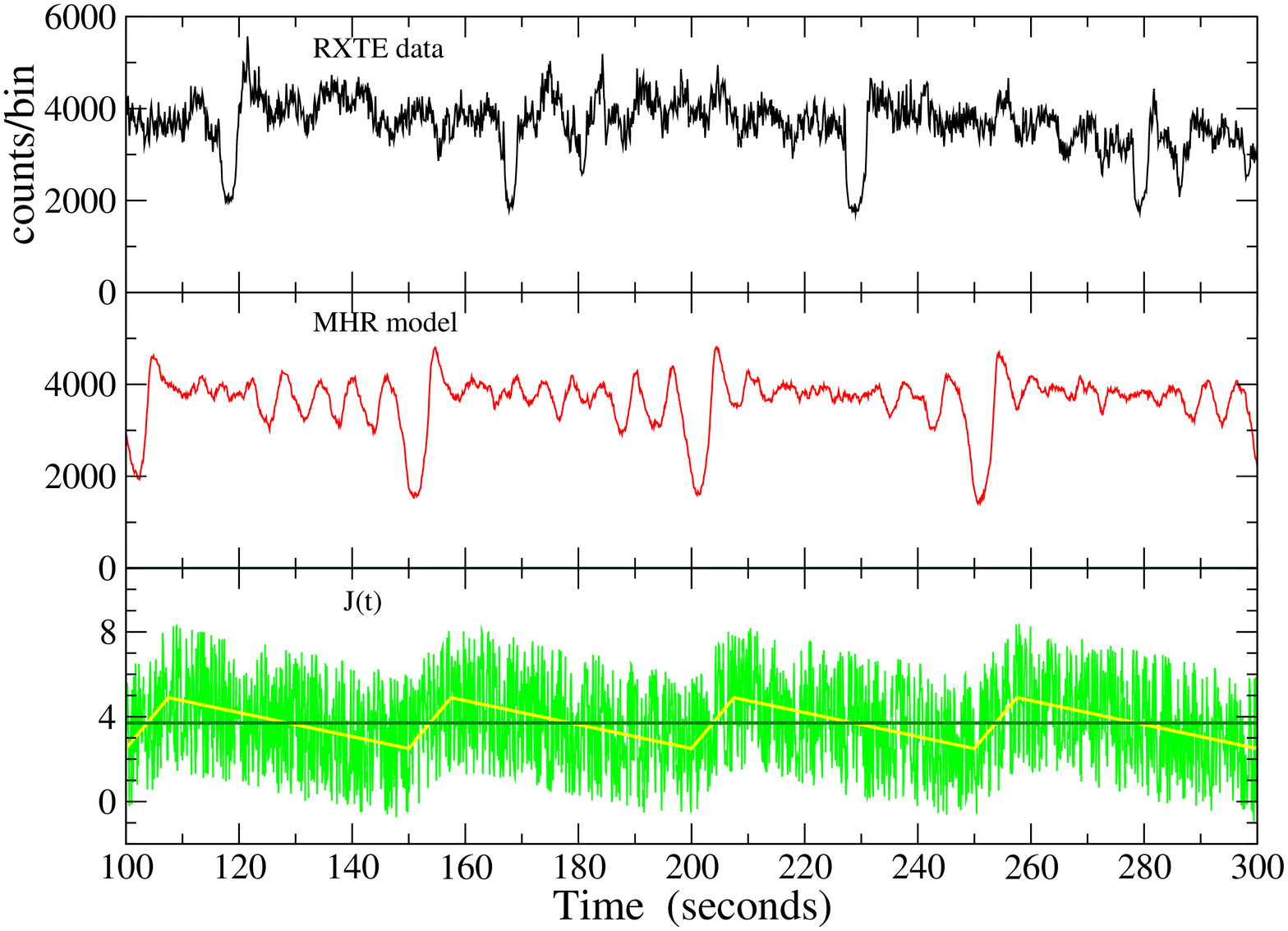}\\
\includegraphics[height=7.9cm,angle=-90,scale=1.0]{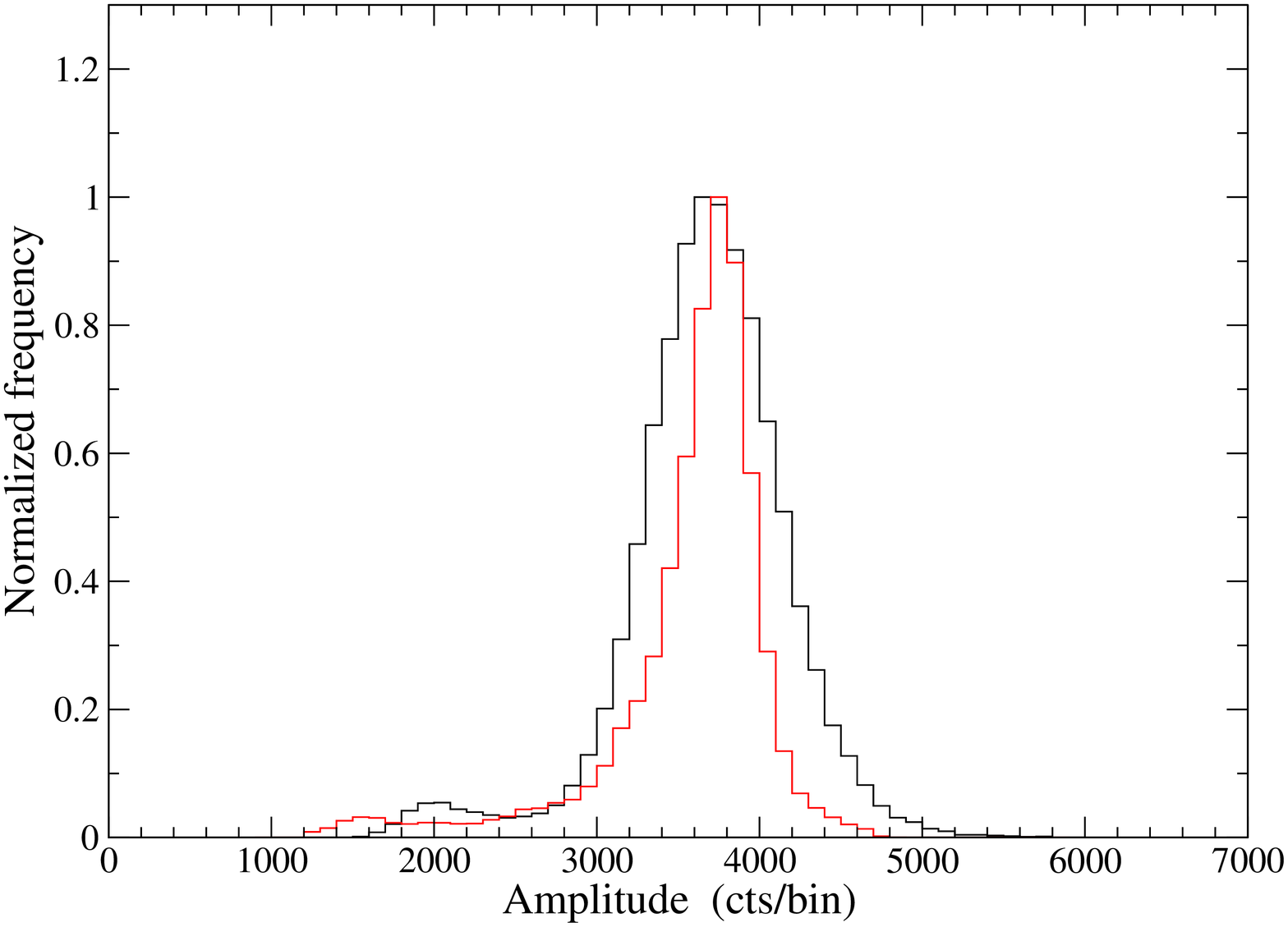}
\caption[]{
Top panel: segment of RXTE light curve of the $\gamma$ class (ID 20402-01-39-00);
centre with a  bin size of 0.125 s; a light curve (red) computed using the MHR 
model using the parameters' values in Table 1; the input function $J(t)$ (green), 
with random fluctuations superposed onto an oscillating function $J_S(t)$ (yellow
curve) with the same time scale of dip recurrence. \\
Central panel: zoom-in of the time series plotted in the top panel to compare their 
structures on short time scales. \\
Bottom panel: histograms of the amplitude distributions in the real data (black) and
computed series (red), normalized to unity at the maxima to compare their asymmetric
profiles.
}
\label{fgamma}
\end{figure}

As shown in the phase space plot (black dashed line in the central panel in Fig.~\ref{falpr}),
in the stable region the trajectory is practically confined in a narrow channel between 
the two {\it nullclines} \citep{Hindmarsh1984, Hindmarsh2005}. 
We plotted here three cubic {\it nullclines}: the central and solid line was computed
assuming $J_0 = \langle J(t) \rangle$, while the other (dashed) two correspond to the 
extreme values of $J_S(t)$, namely 4.2 and $-$0.2.
This portion of the trajectory depends upon the existence of a saddle point between 
the stable and unstable states and explains the shape of the non-spiking segment.
Its evolution is governed by the decay and increasing time scales of $J_S(t)$, while 
the detailed shape is given by the polynomial {\it nullcline} profiles and it is 
remarkably similar between the various bursts (bottom panel in Fig.~\ref{falpr}).
Not, in particular, that its curvature is opposite to that of $J_S(t)$.
Then the trajectory moves into the unstable region and the system changes to the spiking
behaviour that corresponds to the winding loops.
Finally, note that the lowest $x$ points in the loops are very close to the saddle 
point, and this explains the change of direction of the trajectory toward the stable 
channel.

As shown in \citetalias{Massaro2020}, a very interesting property of the $\rho$ class 
spikes is that their recurrence time depends upon the local mean value of $J_S(t)$.
In particular a decreasing $J$ corresponds to an increase of the recurrence time,
and when it becomes lower than the stability threshold the spiking behaviour
is quenched.
A bursting pattern like that of this class was obtained in early works on the original 
HR model (see, for instance, \citealt{Shilnikov2008}) in which an oscillating solution  
for $J(t)$ is obtained from the third ODE in Eq.~\ref{eq1} in  \citetalias{Massaro2020}.
Such a complete system is autonomous and gives also all the solutions described in 
\citetalias{Massaro2020}.
In particular, the equation for the derivative  of $J_S(t)$ depends upon $x$ and $J$, 
but it contains three more parameters.
However, it is hard to obtain solutions with more complex changes and very high gradients 
as those used in the following sections and therefore we preferred to consider this 
function as an independent input.

\subsection{$\beta_2 < \beta_1$: class $\gamma$}

Light curves of the $\gamma$ class are characterized by a rather stable count rate
resembling the $\chi$ or $\delta$ class, with superposed several narrow spikes and 
narrow dips with a time separation of a few tens of seconds (Fig.~\ref{fgamma}, top).
Despite its apparent simplicity these data are not reproduced by means of a constant
$J_S(t)$ and require a more elaborated model with a a rapidly oscillating function 
having the duration of the cycles close to the  separation time between the dips 
(Fig.~\ref{fgamma}, bottom) and a mean value inside the instability range but with 
an amplitude large enough to reach the upper stability domain.
In our calculations we adopted $\beta_1 = 3.0$ as in the previous cases, but fully 
satisfactory results are also obtained adopting values of $\beta_1$ and $\beta_2$ 
lower than 3.0 because, the unstable interval is rather narrow and rather small 
changes of $J_S(t)$ are enough to produce a transition between the two regimes.

A model light curve is given in the top panel of Fig.~\ref{fgamma} (centre), and it 
reproduces many features of the observed curve, in particular the presence of 
recurrent dips.
The model is able to match the data behaviour on time scales as shorth as 10 s,
as shown by the two light curve segments in the central panel in Fig.~\ref{fgamma}.
Note that the amplitude distribution of spikes and dips is asymmetric because the
latter ones have a typical amplitude higher than spikes.
If one computes the histogram of the amplitude values can verify that it is left 
asymmetric; the same property is also present in the histogram of model light curve
(bottom panel in Fig.~\ref{fgamma}), although the properties of the noise are not 
the same as the source.

\subsection{$\beta_2 < \beta_1$: class $\kappa$}

\begin{figure}
\includegraphics[height=7.9cm,angle=-90,scale=1.0]{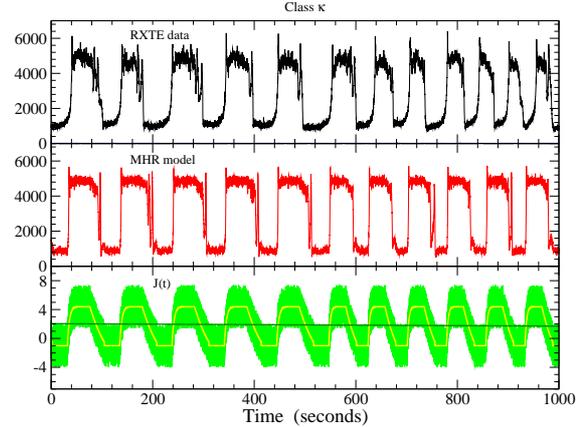}
\caption[]{
Top panel: segment of RXTE light curve of the $\kappa$ class (ID 20402-01-33-00).
Central panel: a light curve (red) computed using the MHR model using the parameters' 
values in Table~\ref{tab:param}.
Bottom panel: the input function $J(t)$ (green), with random fluctuations superposed 
onto an oscillating pattern ($J_S(t)$, yellow line) and its mean value (dark green 
line).
}
\label{fkappa}
\end{figure}

This is a peculiar class because light curves exhibit a large variety of
patterns: also the RXTE observations reported by \citet{Belloni2000} are
largely uneven.
We decided to consider here an observation performed on 1997 June 18
(ID P20402-01-33-00).
The structure of this light curve is characterized by large bursts (their typical
widths range from $\sim$100 s to more than 200 s) with very fast rise and decay; 
narrow spikes frequently occur at the maximum of the rising portion or during the 
decay, as shown in the top panel in Fig.~\ref{fkappa}.
The MHR model works nicely when $\beta_1 > \beta_2$, that in our case are equal 
to 3.0 and 1.6, respectively.
The input function $J(t)$, shown in the bottom panel of Fig.~\ref{fkappa}, is
rather similar to that of the $\gamma$ class with a simple oscillating square 
profile but with a quite lower mean value; the recurrence time of major bursts 
was reduced in the second part of series to make the numerical results more 
similar to the observed data. 
Note that the mean value of these oscillations varies between the stable 
(at the minimum level) and the unstable (at the maximum level) ranges.

The MHR model reproduces not only the typical burst profiles but also some 
details as the fast spikes at the end of the rise and the other one in the 
decay, without any {\it ad hoc} additional feature in the $J_S(t)$ function.
These features are indeed related to the fast transitions segments which can
trigger the onset of a spike mode, soon damped by the fast change of $J_S(t)$.
Clearly, the occurrence of other minor structures in the input function could
produce more complex features in the resulting light curves.

It is interesting that one of the light curves computed by \citet{Janiuk2002}
for a disk model with a radiative instability has $\kappa$-like bursts with 
similar spikes. 
This model was computed for a central black hole of 10 $M_{\odot}$, comparable
to the one estimated for \grs, with about 1/3 of the energy dissipated in the
corona and an outflow, i.e. with additional dissipation processes.
Unfortunately, details of this model and an interpretation of the causes of 
these burstings are not given and a comparison of these finding with ours
is not possible.

\begin{figure}
\includegraphics[height=7.9cm,angle=-90,scale=1.0]{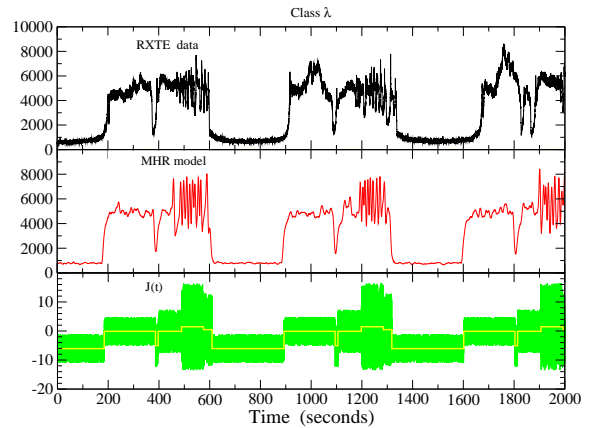}
\caption[]{
Top panel: short segment of RXTE light curve of the $\lambda$ class (ID 20402-01-37-01)
Central panel: a light curve (red) computed using the MHR model using the parameters' 
values in Table~\ref{tab:param}.
The noise is reduced because a running average was applied to this series to have a 
time resolution comparable to those of the data.
Bottom panel: the input function $J(t)$ (green), with random fluctuations superposed 
onto a slowly changing step function with abrupt interruption ($J_S(t)$, yellow curve).
}
\label{flambda}
\end{figure}

\section{Solutions with $\beta_1 = \beta_2$}
\label{sect:5}

Our numerical calculations showed that the simple MHR system proposed in 
\citetalias{Massaro2020} is also able to reproduce the main features of several 
other variability classes, but with the two following assumptions: 
$i$) the values of the input function satisfy the condition that the equilibrium 
point remains in the stable region,
$ii$) the changes of the input function must be over time scales similar to those 
in the observed light curves.

We considered two typical rather simple structures for $J_S(t)$: 
$i$) a series of step functions, repeating the same pattern with a recurrence time 
as the one found in the individual data series; 
$ii$) a combination of slowly modulated variation and step functions, when necessary.
The discussion of a specific physical explanation of these time structures is beyond 
the purpose of this paper. However, assuming they are related to
the local mass accretion rate, this assumption 
requires the occurrence of particular modulations.

Considering that we are working mainly in stable equilibrium conditions we adopted 
$\beta = 3.0$ as in \citetalias{Massaro2020}.
Some values of the parameters of $J_S(t)$ used in numerical calculations are given 
in Table~\ref{tab:param}.

\begin{figure}
\includegraphics[height=7.9cm,angle=-90,scale=1.0]{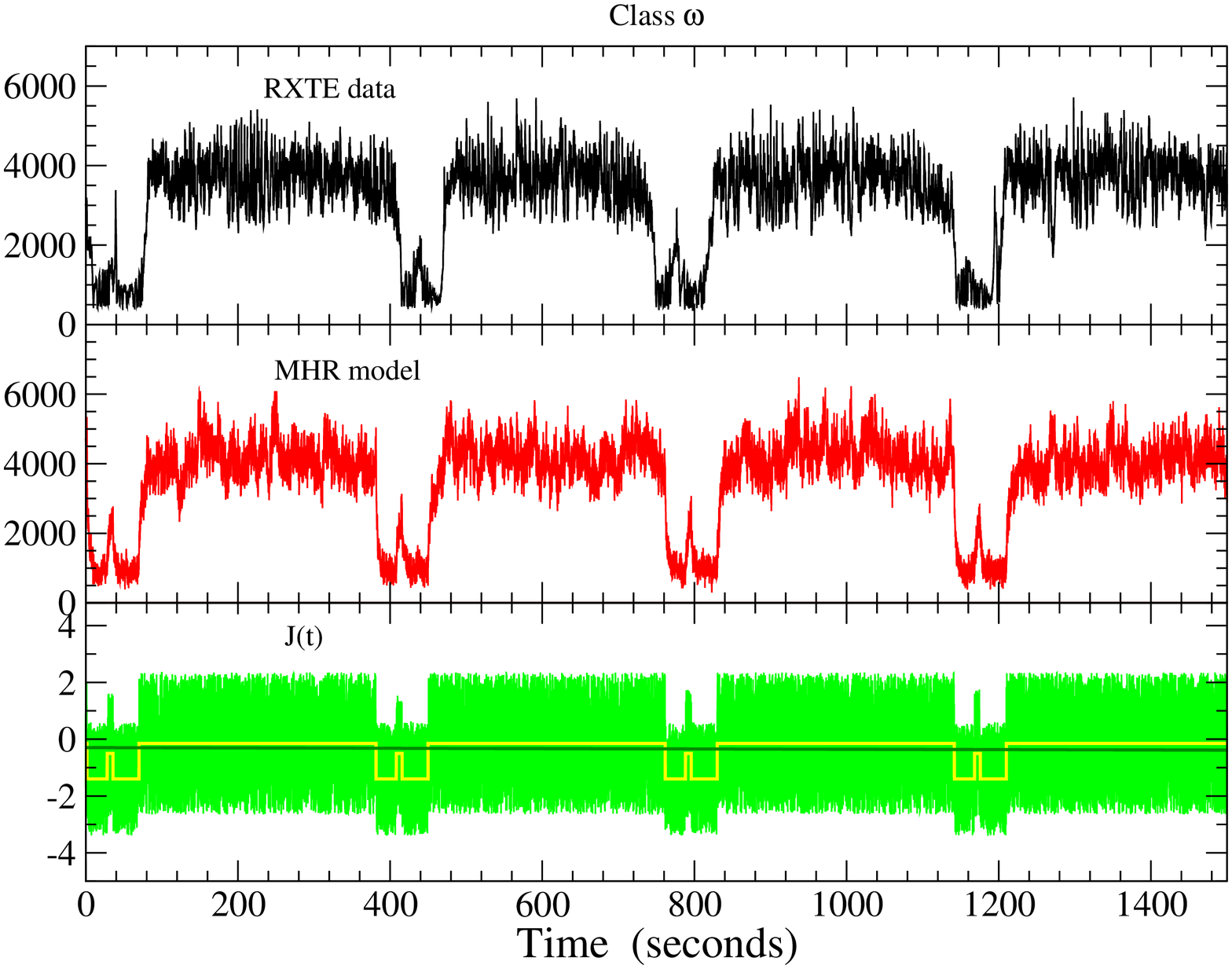}
\caption[]{
Top panel: short segment of RXTE light curve of the $\omega$ class (ID 40703-01-27-00). 
Central panel: a light curve (red) computed using the MHR model using the parameters' 
values in Table~\ref{tab:param}.
Bottom panel: the input function $J(t)$ (green), with random fluctuations superposed 
onto a slowly changing function with abrupt interruption ($J_S(t)$, yellow line).
}
\label{fomega}
\end{figure}

\subsection{Class $\lambda$}
\label{sect:5p1}

This is one of the original classes by \citet{Belloni2000}, characterised by a
complex pattern consisting of rather long bursts with an initial segment similar
to the $\delta$ class followed by a series of fast oscillations.
In the calculation we adopted the choice $\beta_1 = \beta_2 = 4.0$, instead of the
canonical 3.0, but very good solutions are also obtained with $\beta_1 > \beta_2 = 3.0$.
According to the input function models described in Sect.~\ref{sect:3p1}, 
we used a three/four level step function 
(see the bottom panel in Fig.~\ref{flambda}): the lowest level corresponds to the 
minimum steady flux, while of the two higher levels, the one below the zero 
produces the $\delta$ segment and the positive one the fast oscillating, respectively.
A fourth level was introduced to have a smooth decline with an increasing recurrence
time between oscillations.
Moreover, a dip was included in the second segment to obtain a more faithful
correspondence with the observed signal, but this feature is not always necessary
because not present in other $\lambda$ data series.
The resulting light curve is shown in Fig.~\ref{flambda}; 
original results exhibit high amplitude fast oscillations and were smoothed applying
a running average algorithm to make them comparable with data.

\subsection{Class $\omega$}

One of the simplest classes to reproduce is the $\omega$ class, that was recognized by
 \citet{KleinWolt2002} and presents rapidly fluctuations with respect to a rather 
stable level having a duration around 300 s, interrupted by low intensity intervals, 
typically shorter by 80-100 s, and occasionally with a spike in the middle.

We obtained numerical results similar to this class by means of a $J_S(t)$ 
having a three level square wave modulation as shown in the bottom panel in 
Fig.~\ref{fomega}.

\begin{figure}
\includegraphics[height=7.9cm,angle=-90,scale=1.0]{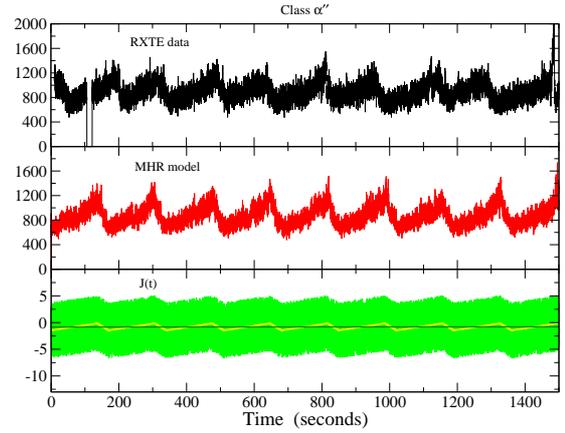}
\caption[]{
Top panel: segment of RXTE light curve of the $\alpha''$ class (ID 20402-01-22-00).
Central panel: a light curve (red) computed using the MHR model using the parameters' 
values in Table~\ref{tab:param}.
Bottom panel: the input function $J(t)$ (green), with random fluctuations superposed 
onto a sawtooth law with $t_1/T = 0.80$; the yellow line is the $J_S(t)$ function, 
and the dark green line corresponds to its mean value.
}
\label{falpha}
\end{figure}

\subsection{Class $\alpha''$}

As writte above, light curves of the \citet{Belloni2000} $\alpha$ class have different 
time structures.
Here we considered the observation ID 20402-01-22-00 with rather smooth variation,
resembling a sawtooth profile, with a recurrence time of about 150 - 200 s, remarkably
different from the $\alpha'$ profile, and that for this reason we report here as
$\alpha''$.
The MHR model reproduces this class by assuming a sawtooth profile for $J_S(t)$ and a 
high statistical noise superposed onto it (see Fig.~\ref{falpha}): note that when the 
$J(t)$ is close to the maxima, a few small amplitude spike can be present, as in the 
real data.

\begin{figure}
\includegraphics[height=7.9cm,angle=-90,scale=1.0]{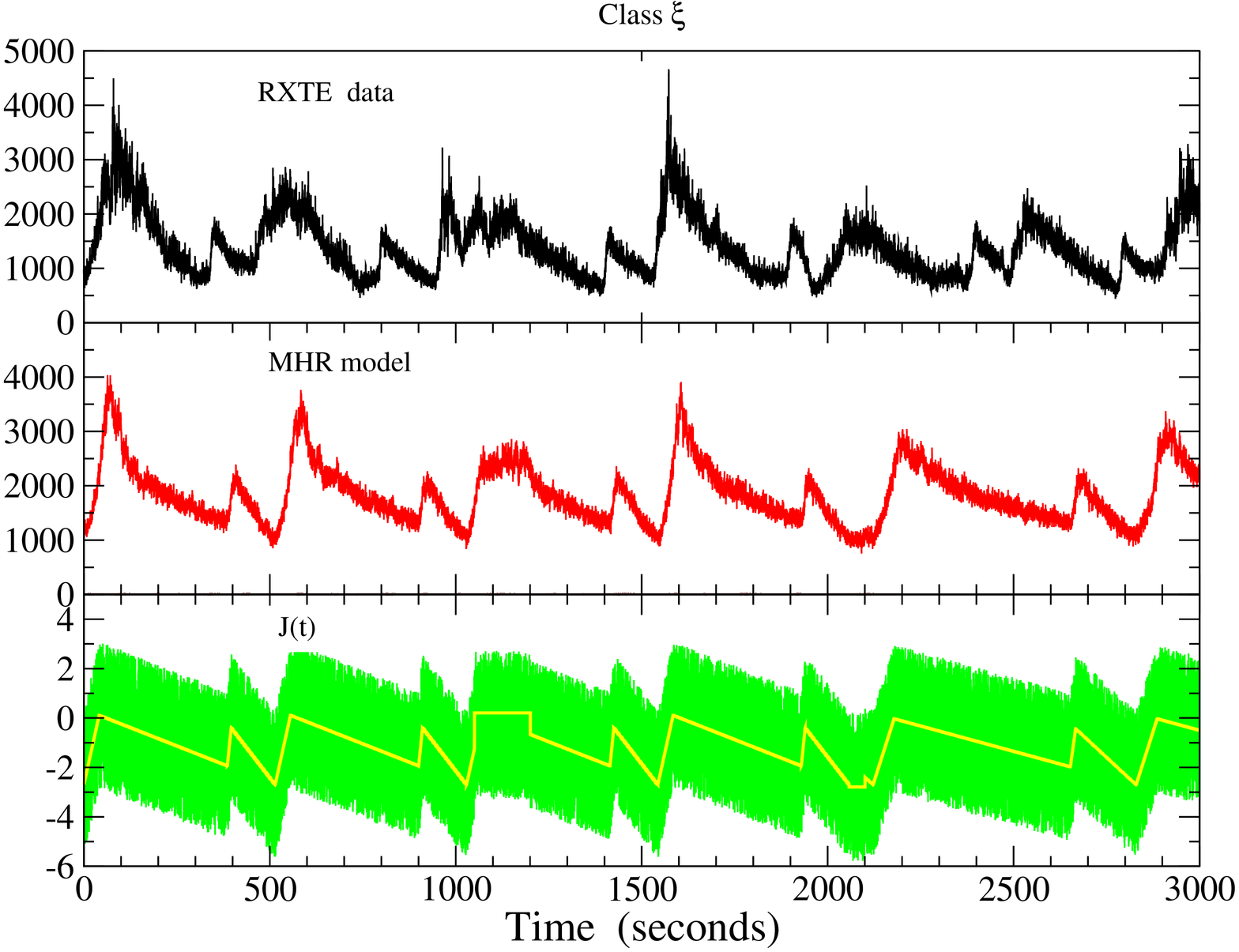}
\caption[]{
Top panel: short segment of RXTE light curve of the $\xi$ class (ID 80127-01-02-01).
Central panel: a light curve (red) computed using the MHR model using the parameters' 
values in Table~\ref{tab:param}.
Bottom panel: the input function $J(t)$ (green), with random fluctuations superposed 
onto a slowly changing function with abrupt interruption, the yellow line corresponds
to $J_S(t)$.
}
\label{fcsi}
\end{figure}

\subsection{Class $\xi$}

This class was firstly reported by \citet{Hannikainen2005}: it consists of a 
triangular modulation characterised by a rise portion shorter than the decaying one,  
at variance with the shape of typical $\rho$ burst.
It looks rather similar to a time reversed $\alpha''$ class.
Duration and height of these triangles are quite variable; the former ones range from
$\sim$100 s to more than $\sim$300 s, and the amplitude ratio can be as high as a factor 
by about 4 (see Fig.~\ref{fcsi}).
We considered a $J_S(t)$ function given by the superposition of two sawtooth of different
amplitudes and added some changes to have a result similar to the observed light curve,
as an increase of the sawtooth duration and a flat segment in correspondence of the
third peak.
What is important is that the mean signal remained below the zero level, that is nearly
coincident to the instability threshold. 
When, in correspondence to the sawtooth maxima, this threshold is occasionally 
exceeded, some narrow spikes are obtained.

\subsection{Class $\theta$}

\begin{figure}
\includegraphics[height=7.9cm,angle=-90,scale=1.0]{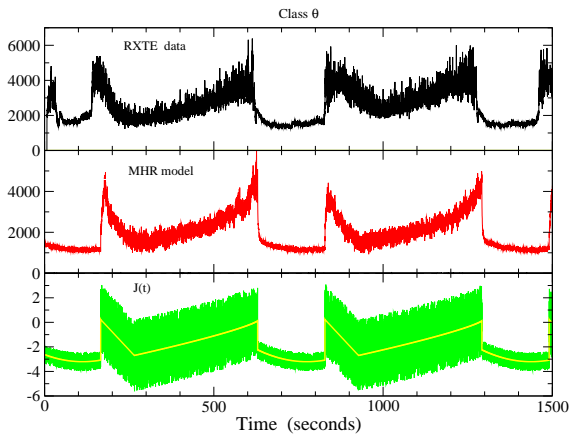}
\caption[]{
Top panel: short segment of RXTE light curve of the $\theta$ class (ID 10408-01-15-04).
Central panel: a light curve (red) computed using the MHR model using the parameters' 
values in Table 1.
Bottom panel: the input function $J(t)$ (green), with random fluctuations superposed 
onto a slowly changing function with abrupt interruption, the yellow line is the $J_S(t)$
the function.
}
\label{ftheta}
\end{figure}

This class was defined by \citet{Belloni2000} and was characterized by cycles over 
a time scale of about 1000 s, with a nearly triangular modulation interrupted around 
the maximum by a low count rate intervals having a duration of 100-300 s and 
occasionally one or more spikes in the middle.
The MHR model reproduces these features if the slow modulation is present in the
$J_S(t)$, as shown in the central and bottom panel in Fig.~\ref{ftheta}.

\section{Summary and discussion}
\label{sect:6}

The results presented in this paper, together with those in \citetalias{Massaro2020},
show that the MHR mathematical model is a useful and powerful approximation
to describe the non-linear processes occurring in the plasma of an accretion 
disc and originating several complex patterns of luminosity variations.
It is, therefore, a promising tool for describing the equilibrium conditions of
the plasma states and for the physical modelling of disc instabilities.
Moreover, the MHR model can help in the understanding of the evolution of
local instabilities originating the various light curve structures.
Furthermore, as shown in \citetalias{Massaro2020}, the model suggests a new
mechanism for the origin of QPOs in the Hz region.

Our calculations have shown that several variability classes of \grss are obtained 
by changing only the input function $J(t)$.
These variations can produce transitions between stable and unstable states, and 
using this property we can group the variability classes into the following three 
main types:

\noindent
$i$) {\it stable equilibrium patterns}: this group includes the classes $\phi$, 
$\chi$, $\alpha''$, $\theta$, $\xi$, and $\omega$, corresponding to variations around
the {\it nullcline} determined by the position of a stable equilibrium with the light 
curve modulated with the same time scale of the changes of $J_S(t)$;

\noindent
$ii$) {\it unstable equilibrium (spiking) patterns}: typically characterized 
by long series of spikes like the $\rho$ (and $\rho_d$) class, it is originated 
when $J(t)$ becomes high enough to move the equilibrium point within the instability
interval; spikes are a recurrence time depending upon the mean level of the 
$J(t)$ and decrease when this value increases;

\noindent
$iii$) {\it transition (bursting) patterns}: this group includes the classes 
$\delta$, $\gamma$, $\lambda$, $\kappa$ and $\alpha'$, in which the variations 
of $J(t)$ produce transitions from stable to unstable regions 
(and {\it vice versa}), thus originating either smooth changes or spiking, 
and occasionally the occurrence of dips and red noise; the $\rho_d$ subclass 
should be classified in this type because of the highly irregular recurrence 
of spikes and the occurrence of plateau between them, which can be due $J$
fluctuations across the stability boundary.

There are other criteria for grouping the variability classes. 
For instance \citet{Misra2004} and \citet{Misra2006}, on the basis of a search 
for a chaotic behaviour, defined the following three groups: non-linear deterministic 
or chaotic classes ($\theta$, $\rho$, $\delta$, $\alpha'$, $\alpha''$), purely 
stochastic classes ($\phi$, $\chi$, $\gamma$), and a mix of stochastic and chaotic 
($\beta$, $\lambda$, $\mu$, $\kappa$).
These groups are different from ours because their classification criteria are
based on the estimate of the correlation dimension and are not very efficient 
for distinguishing the non-stochastic variations in light curves like those of
the $\gamma$ or $\kappa$ classes,

One of the advantages of the MHR model is in its simplicity.
As already pointed out in \citetalias{Massaro2020}, we recall that the model
describes only time changes of the luminosity without considering how these 
depend upon the energy.
The extension of the model, if possible, would likely require at least a new 
non-linear equation and/or the addition of new terms and parameters which would 
make quite difficult the stability analysis of solutions and the conditions for 
the occurrence of a limit cycle. 

\begin{figure}
\includegraphics[height=7.9cm,angle=-90,scale=1.0]{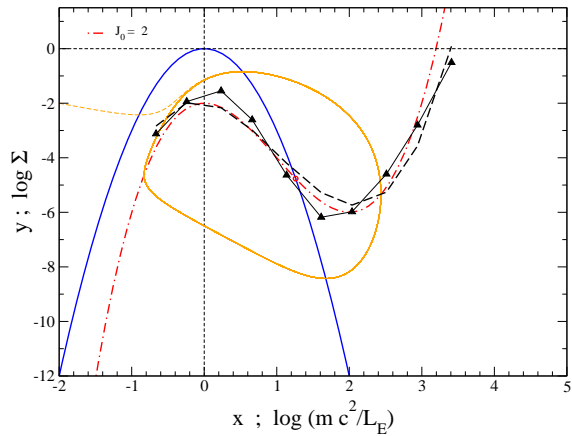}
\caption[]{
{\it Nullclines} of the MHR model of Eqs.(4) for $\beta_1 = 3$ and $J(t) = J_0 = 2$ 
(thick red curve) and the solid blue curve is the parabola for $\beta_2 = 3$.
The red circle marks the unstable equilibrium point.
Black triangles are the values computed by \citet{Watarai2001}
for a slim disk for the logarithms of mass accretion rate in the abscissa and 
that of the disk surface density in the ordinate.
These values were scaled to those of our variables to approximately match the 
curves and dashed black curve is a cubic best fit.
The orange line is the same limit cycle for the $\rho$ class discussed in 
\citetalias{Massaro2020}.
}
\label{fncl_2}
\end{figure}

We now consider a possible physical interpretation of the MHR model and
of the nature of the input function $J(t)$.
In the previous paper concerning the FitzHugh-Nagumo model, \citet{Massaro2014}
proposed that it could be related to the mass accretion rate in the disc.
This hypothesis was suggested by the fact that this 'external' parameter rules the
disc luminosity, but the actual dependence of $J(t)$ from mass accretion rate cannot 
be easily deduced from the data.
It is not, however, the only interesting quantity and here we try to extend the analysis 
to other physical parameters on the basis of the results of some theoretical studies.

One can reasonably rise the question why such complicated dynamical processes are 
captured by a rather simple non-linear ODE system.
There is no simple answer, but a possible indication is that the equilibrium 
conditions for the plasma can be approximated by polinomial functions, at least 
within the limited interval around the unstable states, as those used in the 
MHR model.
This property is found to be common to many physical systems exhibiting 
self oscillations (see \citet{Jenkins2013}, for a clear introduction),
including the macroscopic behaviour of Cepheids variable stars, whose non
linearity is gathered by the period-luminosity relation.
We recall that also for \grss \citet{Massaro2010} reported that the mean recurrence
time of $\rho$ bursts is positevely correlated with the average count rate, 
indicating the occurrence of a similar relation.

Researches on instabilities in accretion disks started in the seventies  
\citep[e.g.][]{Lightman1974, Pringle1973, Shakura1976} 
and up to now an extensive literature has been produced.
\citet{Taam1984}, in particular, computed by means of numerical integration of
the non-linear disk equations, applying the $\alpha$ prescription for the viscosity
\citep{Shakura1973} to investigate thermal-viscous instabilities and 
obtained theoretical light curves having recurrent flares.
With the discovery of the $\rho$ class variability in \grs, \citet{Taam1997}
investigated the time and spectral properties of the bursts and proposed an 
interpretation based on the instability discussed in the previous paper.
The evolution of thermal-viscous instabilities in an accretion disc was associated 
with a limit cycle \citep[e.g. see][]{Szuszkiewicz1998}, generally described by
means of an S-shaped equilibrium curve in a plot of temperature or accretion 
rate vs disk surface density as shown by \citet[][]{Abramowicz1995}.
Other equilibrium curves were computed in a model developed by  
\citet{Watarai2001} and \citet{Mineshige2005} for a slim-disc \citep[]{Abramowicz1988}
as resulting from spectral analyses of \grss \citep{Vierdayanti2010, Mineo2012}.

\citet{Watarai2001} computed the equilibrium relation in the plane 
$log(\dot{m}c^2/L_E), log(\Sigma)$ at a radial distance of 7 Schwarzschild radii, 
where $L_E$ is the Eddington luminosity and $\Sigma$ the surface density of the 
disc, and resulted an S-shaped pattern.
These curves translate along the $log(\Sigma)$ axis with only small changes
of the shape when the viscosity parameter $\alpha$ changes: $log \Sigma$ increases 
by a factor of about 40 when $\alpha$ decreases from 1.0 to 0.01.
The S-profile becomes shallower and shallower for increasing values of the other 
parameter $\mu$, which measures the fractional relevance of the gas pressure to 
the viscous shear tensor: 

\begin{equation}
\tau_{r \phi} = - \alpha P_{gas}^\mu P_{total}^{1-\mu} ~~~, ~~~ 0 \le \mu \le 1.
\end{equation}

\noindent
The structure of these curves is remarkably similar to that of the $\dot{x}$ cubic 
{\it nullcline} of MHR model as apparent in Fig.~\ref{fncl_2} where we reported 
one of the \citet{Watarai2001} curves, precisely the one corresponding to $\mu = 0$ 
and $\alpha = 0.1$, scaled and translated to match the values used in our 
calculations and found a remarkable similarity with the {\it nullcline}. 
For a complete correspondence between these theoretical results and the MHR model 
we need the other relation between $log(\Sigma)$ and $log(\dot{m})$ related to the
$\dot{y}$ ODE and with equilibrium conditions approximating the quadratic 
{\it nullcline} in order to have three equilibrium points necessary for the 
bursting pattern and the interburst profile of the $\alpha'$ class.
It will be useful to verify by means of numerical integrations of slim-disc 
equations if the solutions for physical variables, like the pressure or the temperature 
of the plasma, and consequently the viscous dissipation, follow a similar evolution 
during the limit cycle.

It is also interesting to note that the parameter $\mu$ affects the shape of the 
Watarai \& Mineshige curves and therefore it works as the parameter ratio 
$\beta / \rho$ but in the opposite way.
As for example, assuming a simple linear relationship between these two quantities 
one can write $\beta = k (1 - \mu)$, thus $\mu = 0$ corresponding to the most 
pronounced S-shape (in our case $k = 3.0$). 
The existence of such a relation, however, requires many other calculations to
be verified.

Another interesting possibility to obtain fast changes of $\alpha$ is the one 
proposed by \citep{Potter2014}, who investigated the disc instability induced by 
the dependence of this parameter on the magnetic Prandl number, $Pm$, that is 
the ratio of the plasma (microscopic) viscosity to resistivity.
\citet{Potter2017}, using the PLUTO MHD numerical code \citep{Mignone2007}, 
demonstrated that the $\alpha-Pm$ dependence can be sufficiently strong to produce 
a local instability, which originates a limit cycle by mapping the S-shaped 
thermal-equilibrium curve of the disc.
One can therefore expect that these variations could be the underlying mechanism 
originating $J_S(t)$.

A relevant issue is the definition of the physical meaning of the input function 
$J(t)$ and of the nature of its variations.
In this work we adopted rather simple functions to define the main time scale
useful for computing light curves in a fine similarity to the data, and obtained 
that they were not only in agreement with the general time structure but also 
in several details like, for instance, the spikes of the $\kappa$ class bursts.
The historic light curve of \grss presents luminosity changes higher than 
one order of magnitude \citep{Ghosh2018, Miller2019}, which cannot be explained 
without large changes of an externally driven accretion rate.
A consistent and complete modelling of this activity appears a goal quite hard to 
be achieved, in particular because many important informations are unknown, such 
as the modulation of the gas flow from the companion star.

Finally, we mention that a model for the $\rho$ class was proposed by 
\citet{Neilsen2012}, who interpreted the bursting as a consequence of oscillations 
in the mass accretion rate.
This hypothesis remains useful also in the context of the MHR model, but not,
however, in the specific case of the $\rho$ spikes, but for the $\kappa$ and 
$\gamma$ class that require an oscillating $J_S(t)$ with the same time scale of the 
bursts or dips, respectively.
Using this approach one should search for possible correlations between the 
properties of the various variability classes and the mean luminosity or  
other spectral parameters.
This draining work is beyond the goals of the present paper that is focused on
the development of a tool for describing the stability conditions through a single
model that produce the rich collection of variability classes.

~

\section*{Acknowledgments}
The authors are grateful to Enrico Costa, Marco Salvati and Andrea Tramacere for 
their fruitful comments.
We are also grateful to the referee M. Ortega-Rodriguez for his constructive
comments and suggestions.
MF, TM and FC acknowledge financial contribution from the agreement 
ASI-INAF n.2017-14-H.0.

\bibliographystyle{mnras}
\bibliography{grs1915_II}

\appendix
\section{Nullclines and equilibrium points for $\beta_1 \ne \beta_2$}
\label{appendix1}

To study the solutions when $\beta_1 \ne \beta_2$, we assume that $\beta_1$ is fixed to
3, while $\beta_2$ is variable with the condition to be always positive.
Thus the nullcline for $\dot{x}$ remains equal to that shown in Fig. 1, while the value 
of $\beta_2$ determine the parabolic shape of the $\dot{y}$ nullcline.
The equilibrium points with $J_0 > 0$ are thus given by the solution of the cubic 
equation:
 
\begin{equation}
 x^3 + (\beta_2 - \beta_1) x^2 - J_0 = 0 \nonumber
\end{equation}

For $\beta_2 < 3$ the curvature of the parabola is lower than that in Fig. 1 and there
is only one real solution, while for $\beta_2 > 3$ it is possible to have three real
solutions.
The condition on $\beta_2$ is easily obtained after the reduction of the cubic equation
to the canonical form
\begin{equation}
 u^3 + 3 p u - 2 q = 0 \nonumber
\end{equation}

\noindent
where 

\noindent
 $u = x + (\beta_2 - \beta_1)/3$~~~, ~~$p = -(\beta_2 - \beta_1)^2/9$~~~, \\
$q = (\beta_2 - \beta_1)^3/27 + J_0/2$.

There are three real solutions if the discriminant satisfies the condition:
\begin{equation}
 q^2 + p^3 < 0 \nonumber
\end{equation}
which implies the condition:
\begin{equation}
 \beta_2 $>$ \beta_1 + 3 (J_0/4)^{1/3} ~~,~~ J_0 > 0   \nonumber 
\end{equation}

In the case considered in Section ~\ref{sect:3}, that is $\beta_1 = 3.0$ and 
$J_0 =2.0$, we have three real solutions when 
$\beta_2 > 3 (1 + 2^{-1/3}) = 5.3811...$.

Let $x_*$, $y_*$ the equilibrium solution, the Jacobian computed at this point is:

\begin{displaymath}
\left( \begin{array}{cc}
-3 x_*^2 +2 \beta_1 x_* & 1  \\
-2 \beta_2 x_* & -1  
\end{array} \right)
\end{displaymath}

\noindent
and its determinant and trace are:

\begin{equation}
\Delta = 3 x_*^2 +2 (\beta_2 - \beta_1) x_*   ~~~~,\nonumber
\end{equation}

\begin{equation}
Tr = -3 x_*^2 +2 \beta_1 x_*  -1  ~~~~.\nonumber
\end{equation}

The sign of $\Delta$ depends upon $\beta_1$ and $\beta_2$: for $\beta_1 < \beta_2$ 
we have that $\Delta > 0$ when $x_* < 2(\beta_1 - \beta_2)/3$ or $x_* > 0$, while
for $\beta_1 > \beta_2$ we have that $\Delta > 0$ when $x_* < 0$ or 
$x_* > 2(\beta_1 - \beta_2)/3$.
The trace is $Tr > 0$ only for $(\beta_1 - \sqrt{\beta_1^2 - 3}~)/3 < x_* <
(\beta_1 + \sqrt{\beta_1^2 - 3}~)/3$ and negative outside this interval.

\bsp	 
\label{lastpage}

\end{document}